# A new approach to kinetic energy flux at the different frequencies above the IRIS Bright Points

Rayhane Sadeghi©[1] and Ehsan Tavabi©[1]

[1] *Physics Department, Payame Noor University*
*19395-3697 Tehran, Iran*

## ABSTRACT

Various bright structures abound in the chromosphere playing an essential role in the dynamics and evolution therein. Tentatively identifying the wave characteristics in the outer solar atmosphere helps to understand this layer better. One of the most significant aspects of these characteristics is the wave phase speed (PS), Which is a dominant contribution to solar coronal heating and Energy distribution in the sun's atmosphere layers. To obtain energy flux (EF), it is necessary to calculate the filling factor (FF) and the PS. In this study, the FF was determined by tracking the size and intensity of the *IRIS* bright points (BPs). To estimate an accurate PS and EF, it is necessary to know the chromosphere and transition region(TR) thickness and the phase difference between the two desired levels. chromosphere and TR thickness cannot be measured directly on the disc; This study is performed using spectral data and calibrated based on Doppler velocities. As a result, the PSs in AR and CH, as well as for IRIS BPs have been calculated using the cross-power wavelet transform of Doppler velocities. Consequently, about CH, the PS mean values are from 40 to 180 km/s at network and from 30 to 140 km/s at internetwork; And about AR, are from 80 to 540 km/s at network and 70 to 220 km/s at internetwork. Finally, the EF for the IRIS BPs has been calculated in three different frequencies. The results indicate that the network BPs have an influential role in heating the higher layers while in the internetwork BPs, most of the energy returns to the lower layers.

*Keywords:* Sun: corona — Sun: transition region — Sun: chromosphere

## 1. INTRODUCTION

BPs and activities abound in active regions (ARs), including explosive events, coronal holes (CHs), and quiet Sun (QS) network and internetwork brightness have indicated strong signatures of the magnetic field, and thanks to the unprecedented spatial and spectral resolution of *Interface Region Imaging Spectrometer (IRIS)* (De Pontieu etal. 2014), activities have been seen on even finer scales. chromosphere and TR thickness are very dynamic (e.g. Tavabi et al. 2015b, 2011), and this dynamism makes it difficult to measure the characteristics of this layers (e.g. Leenaarts et al. 2010; Vilinga & Koutchmy 2005). The height of the solar chromosphere is seen in the visible and radio wavelengths. There are problems in measuring this height. For example, solar spicules have different heights. Also, solar flares, usually higher than the chromospheric surface, have variable heights (Aschwanden et al. 2002; Tavabi & Koutchmy 2019; Tavabi et al. 2011). It is difficult to distinguish between the height of the chromosphere and the transition region layer. Measuring the height of the solar atmosphere in the lower layer is more dependent on hydrostatics and energy balance, In contrast, in the upper layers, height values are highly dependent on dynamic factors (Fontenla et al. 2002). In near-solar limb data, photospheric data have the lowest participation rate compared to other spatial locations. As a result, it allows the chromosphere to be examined without disturbing the photosphere data. Some huge values as 28 Mm show the altitudes at the maximum length of macro-spicules, this height more or less refers to the inner corona levels, which is dominated by choromospheric plasma or macro-spicules (Kiss et al. 2017). Using *IRIS* spectral data, Alissandrakis et al. (2018) obtained the altitudes above the visible limb at Mg k & h peaks the 6.909 Mm and 6.539, respectively. The Mg II spectrum has two distinct peaks, k & h (Figure 1), and also smaller peaks in the range between the two apparent peaks (such as single peaks of Fe I, Ni I, Mn I, and Cr II) and so, the most characteristic of them, triple Mg II lines and Fe II lines (Leenaarts et al. 2013; Pereira et al. 2015). These triple lines have wavelengths of 279.160, 279.875, and 279.882 nm. The triple lines are formed due to a sharp rise in temperature of more than 1500k in the lower chromosphere; The Fe II peak has a wavelength of about 280 nm (Leenaarts et al. 2013; Pereira et al. 2015). In addition, Tian et al. (2014) found sub-arcsecond Bright Dots implicating heating events in the TR on ARs above the umbrae and penumbrae of sunspots. Deng et al. (2016) reported that these BDs are unrelated to the chromosphere and suggested their TR formation.

Kleint et al. (2014) presented IRIS evidence of small-scale brightness- which are low corona small loops that emit in UV (Madjarska 2019; Berghmans, D. et al. 2021; Li, Dong 2022)- associated with supersonic down-flows, which have enough energy to heat the TR above sunspots. However, penumbral jets have been considered with IRIS observations and were suggested to be heated to TR temperature (Vissers et al. 2015).

In the past, solar physics researchers have used chromosphere and TR thickness values using ground-based and space observational data and with different element lines (e.g. Auchere et al. 1998; Filippov & Koutchmy 2000; Georgakilas et al. 1999; Johannesson & Zirin 1996). Alissandrakis et al. (2018) in their research on Spicules and Structures Near the Solar Limb, reported that the value of height limb (in the area where there are no coronal holes) for the Mg II k peak is about 9.5 Mm, this value for Mg II h is slightly less than 9 Mm and about 2.5 Mm for Mg II triplet 1 peak (279.160 nm), slightly less than 3 Mm for Mg II triplet 2 peaks (279.875 nm) and about 1.5 Mm for Mg II triplet 3 peaks (279.882 nm) and this amount is slightly less than 1.5 Mm for Fe II peak (280.009 nm). IRIS data allows us to obtain the chromosphere and TR thickness values at high spatial resolution(De Pontieu et al. 2014) and calibrate its values in terms of wavelength, consequently Doppler velocity.

Mein & Mein (1976) have stated, that the chromosphere height based on phase can be derived where its phase is equivalent to the Doppler phase shift. Therefore, the phase difference can be obtained based on the height difference. Mein & Mein (1976) calculated the phase speed to be less than 10 km/s. Morton et al. (2012) has calculated the phase speed of about 48 to 325 km / s for different regions of the QS. Abramov-Maximov et al. (2011) have also stated in their research that 30 to 50 km / s for phase speed.

Another factor that affects chromospheric height and phase speed is the nature of study areas (Sturrock 1964; Solanki & Steiner 1990; Athay 2012; Solanki 2004; Carlsson et al. 2019).

Recently, solar research groups have studied the energy flux of the sun's surface in the chromosphere and corona (e.g. Li et al. 2022; Morton et al. 2012; Petrova et al. 2022; Zeighami et al. 2016). Morton et al. (2012) estimated that the total flux of energy on the surface of the Sun which can reach from the chromosphere to the corona is about $170 \wedge 110$ $W.m^{-2}$ for incompressible transverse motions and $460 \wedge 150$ $W.m^{-2}$ for compressional motions. Petrova et al. (2022) has calculated the energy flux of corona high-frequency oscillations to be $1.9 \wedge 0.68$ $kW.m^{-2}$ for 14-second oscillations and $6.5 \wedge 1.4$ $kW.m^{-2}$ for 30-second oscillations.

## 2. OBSERVATIONS

The data used in this study are divided into two categories: Data used to determine chromosphere and TR thickness, and data used to measure phase velocity. IRIS obtained spectra in near-ultraviolet (NUV), far-ultraviolet 1 (FUV1), and far ultraviolet 2 (FUV2). From 1332 A to 2834 A. Slit jaw images (SJIs) of IRIS by using various filters, can provide images centered on Mg II wing, Mg II k, Si IV 1403 A , and C II ( De Pontieu et al. 2014, and see the IRIS Technical Note 20 for details).

It should be noted that Mg II lines are usually for plasma with low temperatures and above the minimum temperature $\tau_{500} = 1$ because it is an element with the low-First Ionization Potential (FIP). The achieved velocity resolution for IRIS spectra is 0.5 km/s

The observational data used for determine chromosphere and TR thickness were 4 IRIS sit-and-stare data sequences of different Sun's areas ; These data include spectra of QS and AR of the Sun and coronal hole(CH)s area and limb filament loop (LL).

The first set of data relates to the North Pole and coronal holes. The second category is loop limb filament data. The third category is data on the QS in the south pole. The fourth category is the data on the active area of the Sun. A brief overview of the IRIS data (level 2 ) used in this article is given in table 1. The location of the selected data in this section is visible on the left side of the panels a, b, c, and d of figure 2, and are indicated by green rectangles. In the three IRIS data of AR, QS, and CH, the slit is perpendicular to the Sun's surface, but not in the limb filament

**Table 1.** IRIS data

|  | Coronal hole | limb filament loop | Active region | Quiet Sun |
| --- | --- | --- | --- | --- |
| Time [UT] | 2017-08-15 23:10:06 to 00:49:31+1d | 2014-08-17 10:06:13 to 13:59:48 | 2014-08-28 09:06:09 to 10:06:05 | 2014-02-26 00:44:30 to 01:56:18 |
| X, Y | -5",957" | 828",-466" | 997",62" | 6",-981" |
| Max FOV | 119"x119" | 167"x175" | 119"x119" | 119"x119" |
| Roll | 0 deg | -45 deg | 90 deg | 0 deg |
| Raster FOV | 0"x119" | 0"x175" | 0"x119" | 0"x119" |
| Raster steps | 1667x0" | 884x0" | 380x0" | 440x0" |
| Raster step Cad | 3.6 s | 15.9 s | 9.5 s | 9.8 s |
| Raster Cad | 4 s, 1 ras | 16 s, 1 ras | 9 s, 1 ras | 10 s, 1 ras |
| SJI FOV | 119"x119" | 167"x175" | 119"x119" | 119"x119" |
| SJI Cad | 1330:4s, 1666 imgs | Si IV (1400):32s, 441 imgs  Mg II h/k (2796) : 32s, 440 img | 1330:19s, 189 imgs  Si IV (1400) : 19s, 190 imgs | 1330:20s, 220 imgs  Si IV (1400) : 20s, 220 imgs |
| OBSID | 3624255503 | 3800111404 | 3820259253 | 3800009253 |

NOTE—Specifications of 4 series of data used to measure chromosphere thickness, which are sit-and-stare data from *IRIS*. This data is from CH, LL, AR, and QS regions.

**Table 2.** IRIS observations

|  | coronal hole | active region |
| --- | --- | --- |
| Time [UT] | 2016-10-14 20:23:19 to 21:45:03 | 2016-08-02 17:59:15 to 19:58:45 |
| X, Y | 24",-68" | -63",88" |
| Max FOV | 119"x119" | 120"x119" |
| Roll | 0 deg | 0 deg |
| Raster FOV | 0"x119" | 0"x119" |
| Raster steps | 512x0" | 1383x0" |
| Raster step Cad | 9.6 s | 5.2 s |
| Raster Cad | 10s, 1 ras | 5s, 1 ras |
| SJI FOV | 119"x119" | 120"x119" |
| SJI Cad |  |  |
| Si IV (1400): | 10 s, 511 imgs | 16 s, 460 imgs |
| Mg II h/k (2796) : |  | 16 s, 461 imgs |
| Mg II w s (2832) : |  | 16 s, 461 imgs |
| OBSID | 3620259603 | 3620106803 |

NOTE—Specifications of two series of data used for phase speed and energy flux analyzes, which are sit-and-stare data from IRIS. This data is from CH, and AR regions.

loop; Moreover, this difference distinguishes this sample from the other three data. However, there is another great point about this data, which is the existence of a loop in QS areas. The data used for phase speed measurements are the *IRIS* sit-and-stare data possesses and spectrum for the two active, and coronal holes in the center of the solar disk. A brief overview of the IRIS data (level 2 ) used in phase speed and energy flux measurements is summarized in table 2.

3. METHODS AND DATA REDUCTION

Methods and data reduction in this article consists of four parts:chromosphere height, phase speed measurement, filling factor, and energy flux.

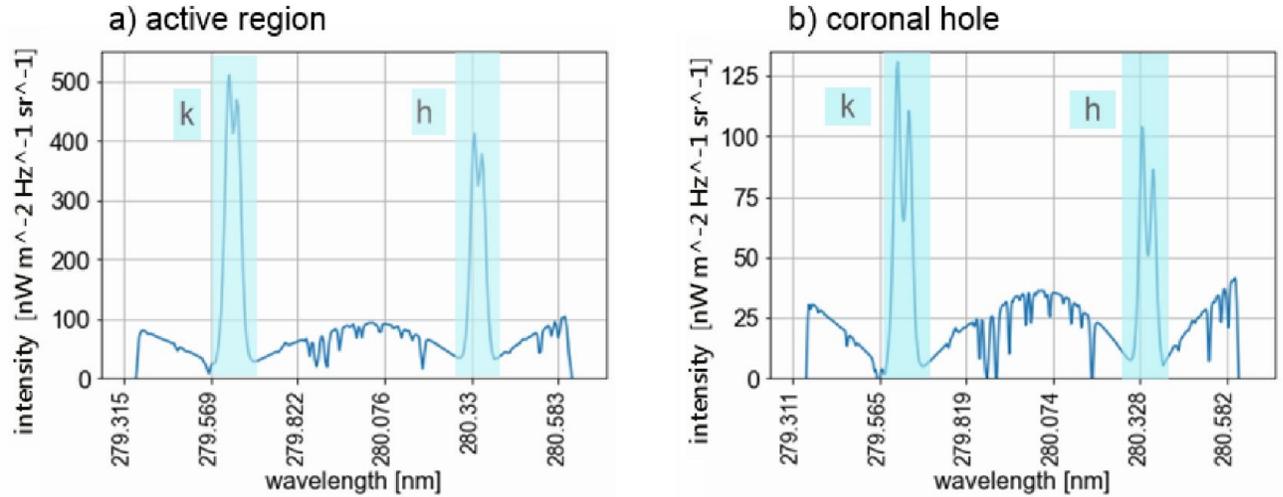

**Figure 1.** a) active region: Mg II distribution diagram of active region (2014-02-25 at 12:56:11 UT ). Investigation of the Mg II spectrum reveals two index peaks,k and h, each peak highlighted by blue. b) coronal hole: Mg II frequency distribution diagram of active region (2014-08-16 at 22:59:47 UT ). Investigation of the Mg II spectrum reveals two index peaks, k and h, each peak highlighted by blue. The valleys of this index peak are identified by the names $k_3$ and $h_3$, which correspond to the peaks k and h, respectively.

First, the IRIS data related to the Mg II spectrum related to specific wavelengths with a width of 0.0025 nm (average data taken from the width of 0.0076 nm) are separated and stacked together based on time, and the intensity time slice corresponding to the specific wavelength is created. Once the intensity time slice is obtained, this data is averaged over time and converted into a column with a width of 0.0025 nm. The chromosphere and TR thickness corresponding to the specified wavelength is then read from the data column. This process is performed during the frequency range of the research, which is finally obtained by plotting the chromosphere and TR thickness in terms of the wavelength, right side of a, b, c, and d panel of figure 2 For QS, AR, CH, and LL, respectively. In the diagrams, five prominent regions are seen, which includes two peaks of the Mg II triplet and two peaks of the Mg II k & h spectrum and the Fe II line core. For Doppler velocity calculations, we need to select a reference as zero velocity. Here, the center of the k peak is chosen as zero velocity, and as we move to the right or the red wing, the velocities increase in a positive direction, and as we move to the left or the blue wing; the velocity values become more negative. Therefore, the zero-velocity referred to the wavelength of 279.639 nm.

To calculate the phase difference of Doppler velocity oscillation, first it is necessary to calculate the Doppler velocity. These calculations are performed using Equation 2. Doppler velocity calculations are performed on the Mg II spectrum for two sets of network and internetwork points in the active and the CH areas. They are calculated for 40 km/s Doppler velocity. For this selection, the maximum Doppler shift graph for Doppler velocity is plotted for Mg II h and k peaks (Figure 3). To perform phase speed analyses, 5 BPs in the active area and 4 BPs in the CH area were selected. These points are selected using SJI, and according to what Sadeghi & Tavabi (2022); Tavabi & Sadeghi (2022) has stated, that based on BPs oscillations period, it is determined which group of network and internetwork they belong to. For this purpose, the time series of the Mg II spectrum in the desired range is analyzed by the Morlet wavelet (Torrence & Compo 1998) and the period of oscillations of the BPs obtained [1]; As a result, network and internetwork points are determined. These BPs are shown in Figure 4 on the SJIs 1401.

Doppler velocity spectra are then given in pairs as input to the Cross-correlation wavelet transform (XWT) function (Grinsted et al. 2004). This function identifies areas with a high common in the frequency and time period, and the phase difference of the input oscillations can be obtained with optimal time and frequency accuracy [2].

To do this, the non-thermal Doppler velocity time slice must first be obtained for the peaks under study (here k3 and h3). Equation 2 then gives the Doppler velocity for the desired speed (here ~ 40 km / s with a tolerance of ^ 10.75 km / s, and slides with a time width of 5.2 seconds for the active area and 9.6 seconds for the coronal hole region, at are put together to create a Doppler velocity time slice. These Doppler velocity time slices are given as cross-violet inputs, and in this way, the phase difference in the desired oscillation periods can be obtained, and thus the phase speed can be calculated with the help of chromosphere and TR thickness. Cross-correlation wavelet transform diagrams are shown in figures 7, 8, 9 , 10 ,and 11 for the AR, and figure 12, 13 , 14 ,and 15 for the CH.

---

[1] The used wavelet analyzes code can be freely downloaded under http:/paos.colorado.edu/research/wavelets/
[2] The used Cross wavelet analyzes code can be freely downloaded under http://grinsted.github.io/wavelet-coherence/

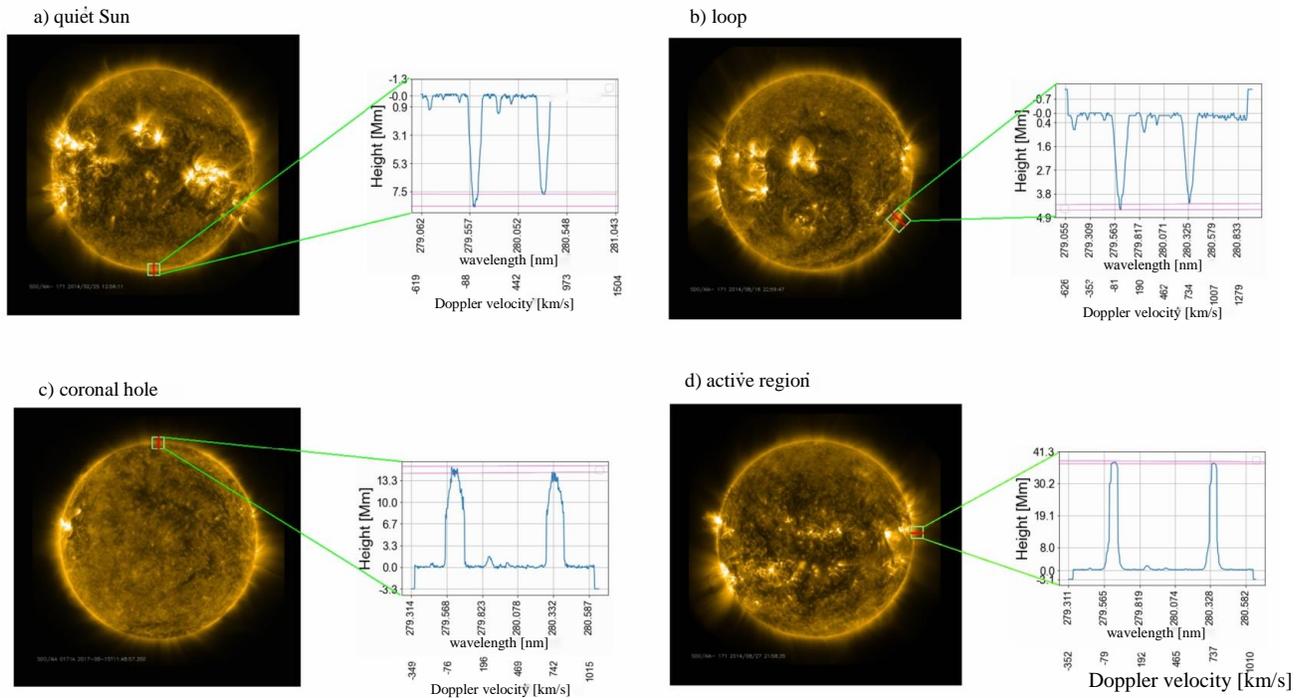

**Figure 2.** a) QS-Left: Solar disk observed in the AIA 171 A channel. The image was taken on 2014-02-25 at 12:56:11 UT by SDO. The green rectangle shows the QS area observed in this study. Right: Graph of chromospheric height in terms of wavelength. The magenta lines indicate the height of the peaks h and k. b) limb filament loop- Left: Solar disk observed in the AIA 171 A channel. The image was taken on 2014-08-16 at 22:59:47 UT by *SDO*. The green rectangle shows the limb filament loop observed in this study. Right: Graph of chromospheric height in terms of wavelength. The magenta lines indicate the height of the peaks h and k. c) coronal hole-Left: Solar disk observed in the AIA 171 A channel. The image was taken on 2017-08-15 at 01:48:57 UT by *SDO*. The green rectangle shows the coronal hole area observed in this study. Right: Graph of chromospheric height in terms of wavelength. The magenta lines indicate the height of the peaks h and k. d) active region-Left: Solar disk observed in the AIA 171 A channel. The image was taken on 2014-08-27 at 21:58:35 UT by *SDO*. The green rectangle shows the active region observed in this study. Right: Graph of chromospheric height in terms of wavelength. The magenta lines indicate the height of the peaks h and k.

The filling factor is one of the factors that can be used to calculate energy flux in different areas. The amount of this component varies depending on the activity of the study area. The filling factor is defined as the ratio of the area of BPs to the total area of the study area (Van Doorsselaere et al. 2014). To obtain this factor, we first divided the BPs into two categories, network and internetwork, according to the dimensions and intensity of BPs; This process was performed on 1400 A images of IRIS and *SDO*/HMI magnetograms at the same time. In figure 6 , the BPs of the network and the internetwork are tracked according to the dimensions and intensity on the SJI Si IV (1400 A), and the BPs of the internetwork are marked with red triangles and the BPs of the network with black circle. Also, with the same method, in figure 6, the internetwork and network BPs on the SDO/HMI magnetogram are marked with red triangles and black circles, respectively. With the help of the specified points and calculating the level of these points and calculating the ratio of the level of these points to the total level, the filling factor can be obtained. The values of obtained for the filling factor in the coronal hole area are given in table 3.

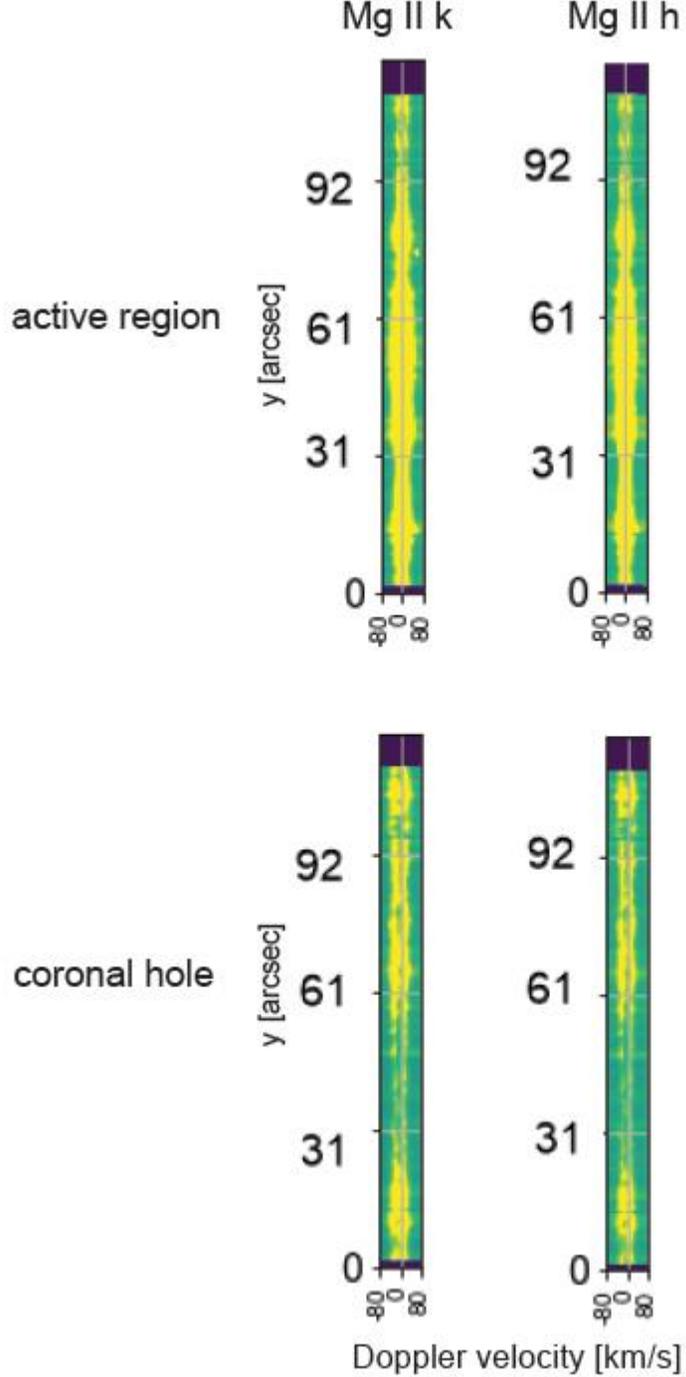

**Figure 3.** The maximum Doppler velocity diagram shows well the red and blue shifts of the spectrum based on the Doppler velocity. Upper left The maximum Doppler velocity diagram of AR-Mg II k. Upper right: The maximum Doppler velocity diagram of AR-Mg II h. bottom left: The maximum Doppler velocity diagram of CH-Mg II k. bottom right: The maximum Doppler velocity diagram of CH-Mg II h.

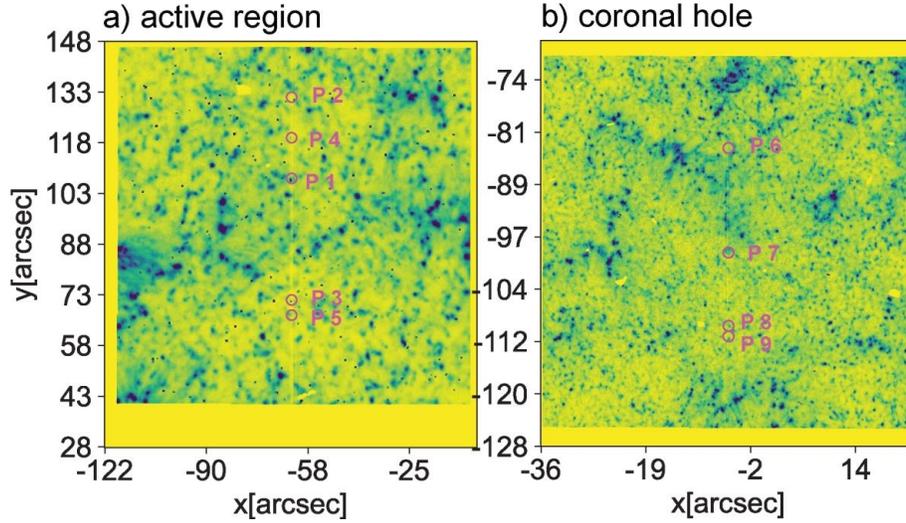

**Figure 4.** a) The positions of points P1 to P5 are shown on SJI 1403 $A$ (Si IV). This SJI is related to the AR, which was captured on 2016-08-02. In this figure, the intensity is inverted and negatived. b) The positions of points P6 to P9 are shown on SJI 1403 $A$ (Si IV). This SJI is related to the coronal hole area, which was captured on 2016-10-14. In this figure, the intensity inverted and negative.

**Table 3.** filling factor of CH

|  | period time | | |
| --- | --- | --- | --- |
|  | 64 s | 180 s | 300 s |
| network | 12% | 2% | 5% |
| internetwork | 12% | 9% | 0% |

NOTE—This table shows the filling factor for the coronal hole area and for periods 64, 180 and 300 s.

The energy flux depends on the phase velocity, Doppler velocity, density of BPs, and filling factor, and formula this table shows the filling factor for the CH and for periods 64, 180 and 300 s (where F is the energy flux, f is the filling factor, p is the BPs density, v is the Doppler velocity and $v_{ph}$ is the phase speed) can be used to calculate the energy flux in different regions of the solar surface (Bate et al. 2022; Van Doorsselaere et al. 2014).

In this way, the energy flux are obtained using the Doppler velocity and the phase speed and filling factor obtained.

$$F \approx f \frac{1}{2} \rho v^2 v_{ph} \qquad (1)$$

4. DISCUSSION AND CONCLUSION

In the present work, we report that chromosphere and TR thickness in the Mg II spectral lines is investigated and its calibration is performed based on the Doppler velocity in different regions of the Sun, including the poles and the equator and near the equator, and then the phase speed is determined in each region (corona hole and active area) and on the two types of chromospheric BPs (network and internetwork).

Determining the energy flux and the contribution of each phenomenon from the energy flux transferred to the higher parts of the solar atmosphere, or returned to the lower levels, is one of the most exciting discussions for solar physics researchers and helps better to understand the solar atmosphere and solar heating mechanism.

For this purpose, it is necessary to calculate the filling factor, which is calculated in section 3.3. filling factor is, in fact, the ratio of the cross-section of BPs to the total area (Van Doorsselaere et al. 2014). This amount has been calculated by Van Doorsselaere et al. (2014) less than 20%. Makita (2003) has also reported a value of 5% by Ca h &k line. Klimchuk (2012) has reported the value of less than 4.5% for the filling factor at QS.

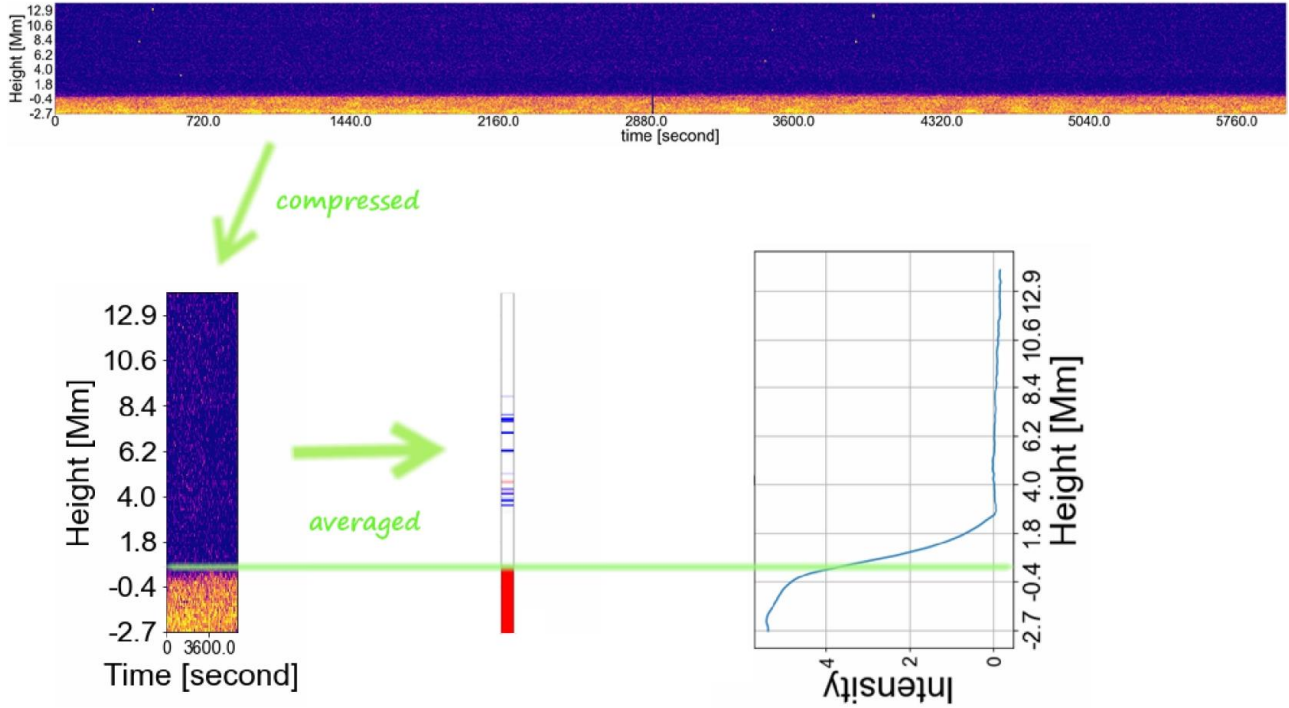

**Figure 5.** Upper: Mg II spectral data for a specific wavelength in terms of time. Bottom-left: Compressed spectral Mg II related to certain wavelength over time. Bottom-middle: averaging spectral data over time and converting it to a column. Bottom-right: intensity-height graphs for accurate readings of specified chromosphere height at each wavelength

Tavabi (2018) has stated that CBPs and BPs are highly interdependent. Hence, their energy is also highly interdependent. pre (1999) has considered CBPs total radiative are about $5.6 \times 10^{27}$ to $1.1 \times 10^{28}$ ergs and conductive energy of CBPs are about $4.0 \times 10^{28}$ to $2.4 \times 10^{29}$ ergs. Also, Priest et al. (2003) have stated that the amount of energy is $^\wedge 10^{18}$ to $10^{20}$ ergs s$^{-1}$.

### 4.1. *line formation height above the limb*

In the chromosphere thickness diagram in terms of frequency in the wavelength range of Mg II, four dominant peaks can be distinguished:
- The first peak of the Mg II triplet peaks is seen around of 279.159 nm.
- The peak of the Mg II k, which is the most apparent peak in this spectral range and is seen around the wavelength of 279.632 nm.
- The peak of Fe II, which is at 280.009 nm, is another peak seen in the Mg II spectral line.

- The other Mg II triplet peaks in the chromosphere and TR thickness diagram are combined and represented as a peak. These two peaks have wavelengths of 279.875 nm and 279.882 nm. However, they can be seen in the chromosphere and TR thickness diagram, around a wavelength of about 279.879 nm.
- The peak of the Mg II h, which is one of the most apparent peaks in this spectral range and is seen around the wavelength of 280.358 nm. chromosphere and TR thickness values are calculated for 4 regions and selected times from the Sun, and the five index peaks mentioned (Table 4).

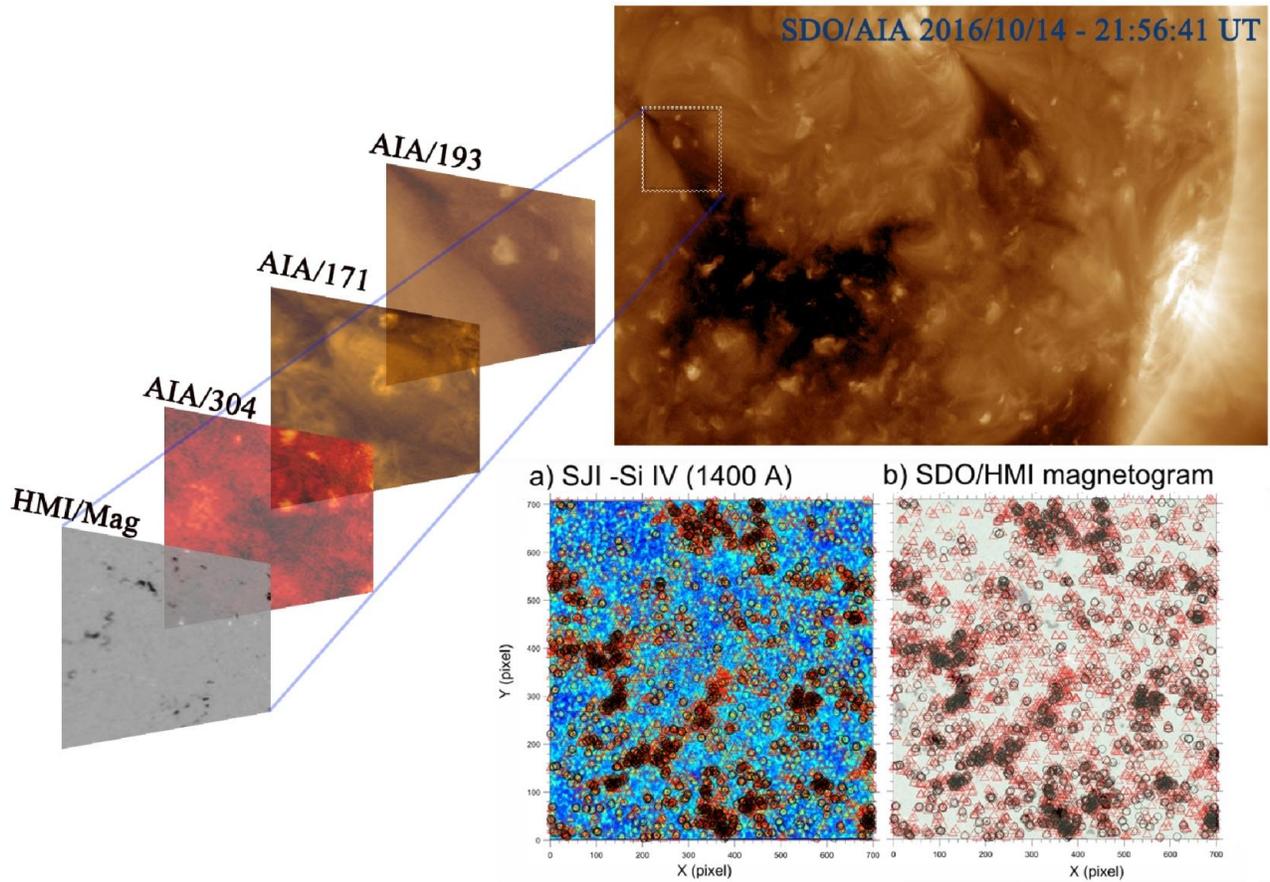

**Figure 6.** SDO/HMI images in three wavelengths of 193, 171 and 304 A and SDO/HMI magnetogram related to the photosphere, respectively. a) SJI Si IV(1400 A), in which the network BPs are marked with black circles and the internetwork BPs with red triangles are marked. b)SDO/HMI magnetograms, which the network BPs are marked with black circles and the internetwork BPs with red triangles are marked.

**Table 4.** chromosphere thickness

| Name | Wavelength (nm) | CH (Mm) | LL (Mm) | AR (Mm) | QS (Mm) |
|---|---|---|---|---|---|
| Mg II triplet | 279.159 | 0.7 | 0.8 | 0.5 | 1.1 |
| Mg II k | 279.632 | 15.3 | 4.5 | 39.2 * | 8.8 |
| Mg II triplet 2 | 279.879 | 1.8 | 0.9 | 0.9 | 1.4 |
| Fe II | 280.009 | 0.8 | 0.6 | 0.6 | 0.7 |
| Mg II h | 280.358 | 14.6 | 4.2 | 38.3 * | 7.8 |

NOTE—This table shows the chromosphere and TR thickness of 5 peaks of the Mg II h & k spectrum index for the four regions under discussion(CH, LL, AR, QS). Values marked with an asterisk in the table are more related to the extended chromosphere to the inner corona and are caused by TR penetrated materials.

The height difference between the h and k peaks in each of the four study areas is shown in table 5. Also, the difference between the triplet peaks is expressed in this table.

By comparing the chromosphere thickness values in the QS and the AR, it can be concluded that the chromosphere and TR thickness of the active sun is 5 times larger than that of the QS (this amount is so significant). This means

193

**Table 5.** maximum height difference

| Name | CH (Mm) | LL (Mm) | AR (Mm) | QS (Mm) |
|---|---|---|---|---|
| height difference between the h and k peaks (Mm) | 0.7 | 0.3 | 0.9 | 1.0 |
| height difference between triplet peaks (Mm) | 1.1 | 0.1 | 0.4 | 0.3 |

NOTE—This table shows the chromosphere and TR height differences of 5 peaks of the Mg II h & k spectrum index for the four regions under discussion(CH, LL, AR, QS).

that knowledge of the area under study is essential to determine the chromosphere and TR height.

Also, as can be seen from the height diagrams in terms of wavelength, the height values change significantly with the wavelength changes. These height in terms of wavelength can be directly related to changes in chromosphere thickness in terms of changes in all physical parameters associated with wavelength.

Some studies have shown that number of spicules can transfer a lot of mass and energy to the corona (e.g. Tavabi et al. 2015a). In active regions, magnetic structures higher than other regions are observed. In polar regions, accumulation of coronal hole jets is seen due to open magnetic fields. The altitudes above the 10 Mm beyond to inner corona, however in huge prominence and in active regions with TR mini-loops and arcs or monofilaments, the chromospheric rather cool material emission lines are penetrated and seen in higher altitudes, even the low-FIP photospheric element could reach to the inner corona and freely contributed to the fast solar wind. Values marked with the asterisk in the table 4 are more related to the extended chromosphere to the inner corona and are caused by TR penetrated materials (Bennett & Erdelyi 2015; Sow Mondal et al. 2022; Martinez-Sykora et al. 2018).

$$\Delta v_{Doppler} = \frac{-1}{2} \frac{c}{\lambda_{k_3}} [(\lambda_{k_{2v}} - \lambda_{k_3}) + (\lambda_{k_{2r}} - \lambda_{k_3})] \quad (2)$$

where c, $A_{k_3}$, $A_{k_{2v}}$, $A_{k_{2r}}$ are respectively, speed of light, $k_3$ line centre wavelength, $k_{2v}$ observed $k_v$-peak wavelength, $k_{2r}$ observed $k_r$-peak wavelength.
As a result, by considering a reference point for the Doppler velocity, each wavelength can be assigned a number commensurate with the Doppler velocity, and the chromosphere thickness diagram can be plotted against the Doppler velocity.

### 4.2. *phase speed measurement*

Zeighami et al. (2020) studied the Doppler velocity above chromospheric network and internetwork points and obtained values of -21 to +21 for internetwork points and -20 to 30 km / s for network points. Zhang et al. (2021) reported the Doppler velocity value of flare-related coronal jet, about -120 to 170 km/s.
We have recorded the maximum Doppler velocity observed for the AR and CH Mg II h & k as shown in table 8 (as shown in figure 3). The maximum Doppler shift velocity at the network BPs is - 49 km/s for the blue shift and +50 km/s for the red shift, at the active region. The maximum phase speed at the internetwork BPs is -33 km/s for the blue shift and +30 km/s for the red shift at the active region. These values for the coronal hole regions are -45 km/s for the blue shift and +35 km/s for the red shift, at network BPs; And -30 km/s for the blue shift and +25 km/s for the red shift, at internetwork BPs. In general, it can be said that the shifts related to network BPs are more than similar BPs at internetwork and the number of shifts in k-peaks is more than the corresponding h peak.

As a result, for the analysis of this article, 40 km/s has been selected as a value of Doppler velocity (LOS). Kim et al. (2008) in 2008 stated the phase speed was estimated at 260 to 460 km/s. In 2009, the phase speed was estimated at 50 to 150 km/s by He et al. (2009). Abramov-Maximov et al. (2011) calculated this value, about 30 to 50 km/s and

**Table 6.** phase speed

| BPs area | point category | | period time second | upward phase speed km/s | downward phase speed km/s |
|---|---|---|---|---|---|
| active area | P1 | network | 64 | 180 | - |
| | | | 180 | 790 | -590 |
| | | | 300 | 110 | - |
| | P2 | internetwork | 64 | 100 | - |
| | | | 64 | 40 | - |
| | | | 180 | - | -280 |
| | | | 180 | - | -400 |
| | | | 300 | - | - |
| | P3 | network | 64 | 70 | - |
| | | | 180 | 290 | -60 |
| | | | 300 | 200 | - |
| | | | 300 | 200 | - |
| | P4 | internetwork | 64 | 60 | - |
| | | | 180 | - | -70 |
| | | | 180 | - | -150 |
| | | | 300 | 150 | - |
| | P5 | network | 64 | 30 | - |
| | | | 64 | 30 | - |
| | | | 180 | - | - |
| | | | 300 | 210 | - |
| | | | 300 | 100 | - |
| | | | 300 | 200 | - |
| coronal hole | P6 | network | 64 | 90 | - |
| | | | 180 | - | -170 |
| | | | 180 | - | -90 |
| | | | 300 | 40 | - |
| | | | 300 | 20 | - |
| | P7 | network | 64 | - | -40 |
| | | | 64 | - | -40 |
| | | | 64 | - | -40 |
| | | | 180 | - | -60 |
| | | | 180 | - | -60 |
| | | | 180 | - | -70 |
| | | | 180 | - | -30 |
| | | | 300 | 500 | - |
| | | | 300 | 170 | - |
| | P8 | internetwork | 64 | - | -150 |
| | | | 180 | 350 | - |
| | | | 180 | 190 | - |
| | | | 300 | 30 | - |
| | P9 | internetwork | 64 | - | - |
| | | | 74 | - | -130 |
| | | | 180 | 90 | - |
| | | | 180 | 50 | - |
| | | | 180 | 30 | - |
| | | | 300 | - | - |

NOTE—The calculated phase speed for the BPs of the QS and AR are given in this table. BPs, P 1 to P 5 are related to AR and P 6 to P 9 are related to CH. All calculations are performed for two groups of time periods and several time intervals with high accuracy. The negative phase difference values mean lower currents and positive values related to upward currents.

12           R. SADEGHI & E. TAVABI

**Table 7.** average phase speed

| | | period time second | upward phase speed km/s | downward phase speed km/s |
|---|---|---|---|---|
| active area | network | 64 | 80 | - |
| | | 180 | 540 | -320 |
| | | 300 | 160 | - |
| | internetwork | 64 | 70 | - |
| | | 180 | - | -220 |
| | | 300 | 150 | - |
| coronal hole | network | 64 | 90 | -40 |
| | | 180 | - | -80 |
| | | 300 | 180 | - |
| | internetwork | 64 | - | -140 |
| | | 180 | 140 | - |
| | | 300 | 30 | - |

NOTE—average phase speed of coronal hole and active area for network and internetwork. negative value indicated downward flows and positive value related to upward flows.

**Table 8.** maximum Doppler shift velocity of active area's BPs

| BPs area | BPs | point category | blue shift k peak km/s | red shift k peak km/s | blue shift h shift km/s | red shift h shift km/s |
|---|---|---|---|---|---|---|
| active | P1 | network | -45 | +40 | -42 | +32 |
| | P2 | internetwork | -33 | +30 | -23 | +23 |
| | P3 | network | -47 | +42 | -42 | +38 |
| | P4 | internetwork | -27 | +25 | -23 | +20 |
| | P5 | network | -49 | +50 | -45 | +45 |
| coronal hole | P6 | network | -30 | +30 | -31 | +30 |
| | P7 | network | -45 | +35 | -40 | +32 |
| | P8 | internetwork | -22 | +20 | -22 | +20 |
| | P9 | internetwork | -30 | +25 | -24 | +23 |

NOTE—The maximum observed Doppler shift velocities of the Mg II k & h peaks for BPs in the CH and AR regions are given in this table. Points pi to p5 are related to AR and points p6 to p9 are related to CH.

in 2012, Morton et al. (2012) obtained 48 to 325 km/s for phase speed.

The results of cross-wavelet analysis and phase speed of active region and coronal hole area, respectively are shown in table 6, for the Doppler velocity of 40 km/s at Mg II k & h, and using the height difference obtained for the chromosphere and TR thickness at these two peaks in the previous step.

According to the results, the phase speed is from 30 to 800 km/s in the active region and from 20 to 500 km/s in the CH.

Phase speeds are from 20 to 500 km/s in network BPs of coronal hole area and from 30 to 300 km/s at internetwork BPs of coronal hole area; and Phase speeds are from 30 to 800 km/s at network BPs of the active region and 40 to 400 km/s at internetwork BPs of the active region.

In most of the internetwork BPs with intensity oscillation of the 180 seconds, no high correlation was found in the Doppler velocity of k & h peaks in the period of 300 seconds. Moreover, this is well seen in the results of the coronal holes.

In most points, the phase speed in the 180 second period is faster than in the 300 seconds period, but this is not the case everywhere (ex: point P7).

**Table 9.** Energy flux

| BPs area | point category | upward | | | | downward | | | | net flux |
|---|---|---|---|---|---|---|---|---|---|---|
| | | 64 s W/m² | 180 s W/m² | 300 s W/m² | total W/m² | 64 s W/m² | 180 s W/m² | 300 s W/m² | total W/m² | W/m² |
| coronal hole | network | 5520 (57%) | – (0%) | 4410 (43%) | 9930 | 1920 (75%) | 630 (25%) | – (0%) | 2550 | 7380 |
| | internetwork | – (0%) | 5110 (100%) | 0 (0%) | 5110 | 6670 (100%) | – (0%) | – (0%) | 6670 | -1560 |

NOTE—In this table, the energy flux for the two areas of network and internetwork is calculated from the CH. For these measurements, the average phase velocity in each region is used in two types of upward and downward. Negative values mean recurrent energy to the lower layers and positive values mean the transfer of energy to the higher layers.

In general, it can be said that the phase speed in the active region is higher than in the coronal hole area. Also, in general, the phase speed at the network BPs is higher than in the internetwork BPs. In general, it can be said that the phase speed in the active region is higher than the coronal hole area. Also, in general, the phase speed at the network BPs is higher than the internetwork BPs.

### 4.3. *filling factor*

The filling factor is one of the most critical factors in calculating energy flux. According to what can be seen in figure 6, and by separating the BPs of the network and the internetwork, the filling factor calculations have been performed and the filling factor values for the coronal hole area are given in table 3. It should be noted that the filling factor values for the active area have not been calculated due to the complexity of situations in the active areas. Considering the filling factor values, it seems that in both network and internet areas, the most effective coefficient is seen in high frequency oscillations. After high frequency oscillations, in network areas, 180 seconds oscillations have the highest value and effect, and in internetwork areas, 300 seconds oscillatons have the most effect.

### 4.4. *energy flux*

Equation 1 has been used to calculate the energy flux. In this formula, the average phase speed mentioned in Table 7 and the filling factor values corresponding to the desired area according to Table 3 have been used, and Table 9 has been obtained in which the energy values for the BPs of the network and internetwork at CH.

According to the calculated amounts for the energy flux, at the network BPs:
The total upward energy flux is greater than the sum of the downward energy flux.
The largest share of upward energy flux is related to high frequency oscillations, followed by 300 s oscillations.
The largest share of the downward energy flux is related to high frequency oscillations, followed by 180 s oscillations. In general, it could be said that 300 s oscillations have the largest share in upward energy transfer, followed by high-frequency oscillations.
180 s oscillations contribute the most to the energy return to the lower layers.

At internetwork BPs:

The total downward energy flux is greater than the total upward energy flux.
The most enormous share of upward energy flux is associated with 180 s oscillations.
The most considerable share of downward energy flux is related to high frequency oscillations.
In general, it can be said that 180 s oscillations have the most significant share in upward energy transfer.
High frequency oscillations have the most immense share in the return of energy to the lower layers.

## 5. SUMMARY AND RESULTS

The propagation of disturbances along the axis of BPs can be deduced when observations are performed at different heights in TR. Then the phase speed can be estimated from the phase difference (vice versa) between Doppler velocity oscillations at different heights. Papushev & Salakhutdinov (1994) found the phase differences of fluctuations at different heights and then concluded that the propagation speeds should be larger than 300 km/s. De Pontieu et al. (2007) used *Hinode*/SOT observations and has stated that some of partially standing waves have upward and downward propagation behavior with phase speed of 50-200 km/s.
Gosic et al. (2018) illustrated that fine BPs internetwork magnetic fields play an essential role as a heating agent They gave shreds of evidence of IN magnetic elements cancellations in the photosphere that can make transient brightness in the chromosphere and transition region. These bright structures might be the signature of energy release and heating, probably driven by the magnetic reconnection of internetwork field lines.

In this research, one of the most critical wave characteristics, namely phase speed, in the chromosphere is investigated. For this purpose, the chromosphere and TR thickness was calculated using the Mg II spectrum and the phase speed between the two peaks Mg II k & h was calculated. In this paper, the phase speed of the network and internetwork BPs, at the AR and the CH, has been studied separately. According to the results, the phase speed in the network BPs is higher than the internetwork ones and it also seems that the phase speed at the active areas is more than.

We suggest that the Doppler shift of the central lines depression correlates strongly with the vertical velocity, which is typically placed in order of 0.5 Mm below the TR. By combining the Doppler shifts of the Mg II k & h lines we can retrieve the sign of the velocity gradient just below the TR. Leenaarts et al. (2013) found that the central line intensity and the structure height are anti-correlated with each other, and this anti-correlation is more divulged in few square Mm. This intensity can be used to measure the changes of TR height. The peak intensity of the emission lines has a high correlation with the temperature of the formation height. As a result, these peaks are characteristic for temperature detection and the velocity gradient in the upper chromosphere that is related to the difference in the wavelength of the blue and red peaks. So Mg II k & h lines are excellent markers for upper chromosphere and below TR, this method is only possible for Mg lines. Also, these lines are suitable for detecting temperature and velocity in the middle chromosphere. For the reasons mentioned, many investigations related to the velocities and temperatures of the chromosphere have been done using Mg lines (e.g. Tavabi et al. 2022; Sadeghi & Tavabi 2022).

Zaqarashvili et al. (2010) found that the propagation of the actual oscillations is rather challenging to detect with the relative Fourier phase differences between oscillations at different levels indicating the propagation speed of >> 110 km/s. In some oscillations, which could be related to standing patterns, waves seem to be at higher phase speeds (> 300 km/s).

In conclusion, in this work, we analyzed spectral observations of Mg II line emissions, for the first time at sub-arcsecond resolution accessible with *IRIS*, particularly focusing on the Doppler shifts of the emission of different regions and showing broad distributions of Doppler shifts. These characteristics are primarily confined to the transition region and show shifts caused by episodic heating in the lower solar atmosphere. The viability of the phase speed of wave as a heating mechanism relies upon the efficient dissipation and thermalization of the wave energy, with direct evidence remaining elusive until now. Here we provide and implemented the first observational evidence of phase speed heating TR.

Finally, we suggest that the Doppler velocities oscillations propagate the longitudinal wave in the stratified TR along the axis of magnetic BPs (Sadeghi & Tavabi 2022), which are excited by p-modes buffeting in network and internetwork positions in CH, AR,

- The mean Doppler velocities of network BPs undergo oscillations with periods of 180 s and 300 s at CH and the phase speed is about 130 km/s; - Doppler velocities of internetwork BPs undergo oscillations with periods of 180 s and 300 s at CH and the phase speed is about 80 km/s; - Doppler velocities of network BPs undergo oscillations with periods of 180 s and 300 s at AR and the phase speed is about 340 km/s; - Doppler velocities of internetwork BPs undergo oscillations with periods of 180 s and 300 s at AR and the phase speed is about 180 km/s;

The Doppler velocity waves propagation carries almost all of the energy of initial perturbations. In contrast, the energy in wake oscillations is much smaller than other ones. Therefore, the energy carried into the TR and corona by longitudinal waves with the remarked phase speeds can be much higher than is estimated by observed oscillations. More observations and numerical/analytical works are needed to look further into this problem. It seems that the estimated energy flux of this mode provides the energy required to heat the solar corona, taking into account the dissipated.

The energy flux estimation gives us much information about how energy is transferred between the solar atmosphere's layers and how they heat up. To calculate this critical component, it is necessary to obtain the filling factor in the study area. In this study, we obtained the filling factor values for the CH region and oscillations of 64 seconds (high frequencies), 180 seconds, and 300 seconds. Using the values of the filling factor, the contribution of each of the oscillations in the energy supply of the higher layers has been determined.

- The mean Doppler velocities of network BPs undergo oscillations with periods of 180 s and 300 s and high frequencies at CH and the net energy flux is about 7380 $W/m^2$;
- Doppler velocities of internetwork BPs undergo oscillations with periods of 180 s and 300 s at CH and the net energy flux is about -1560 $W/m^2$ (Negative values mean the return of energy to the lower layers and positive values mean the transfer of energy to the higher layers).

An important point to note at the end is that all calculations have been made for energy flux at Doppler velocities of 40 km/s, and by calculating the share of other velocities it can increase the net flux, and this increase probably has a value about below 20%.

Finally, it should be emphasized that the typical amount for energy flux at a Doppler velocity has been calculated of the dominant Doppler peak around 40 km/s in TR, and this value must be integrated all over the other velocities; It can significantly increase the energy net flux.

## ACKNOWLEDGMENTS

We thank the referee for her/his valuable comments, which have improved the paper. We acknowledge IRIS for the publicly accessible data used in this paper. IRIS is a NASA small explorer mission developed and operated by LMSAL.

The AIA and HMI data used here are provided by *SDO* (NASA) and the AIA and HMI consortium.

Cross Wavelet and Wavelet Coherence Toolbox were provided by A. Grinsted, J. C. Moore and S. Jevrejeva, and is available at http://grinsted.github.io/wavelet-coherence/

Wavelet software was provided by C. Torrence and G. Compo, and is available at http://paos.colorado.edu/research/ wavelets/

DATA AVAILABILITY

The IRIS data which were used in this article are available at https://iris.lmsal.com The AIA and HMI data of SDO are publicly available at http://jsoc.stanford.edu/

Wavelet software which were used for wavelet analysis is available at http://paos.colorado.edu/research/wavelets/

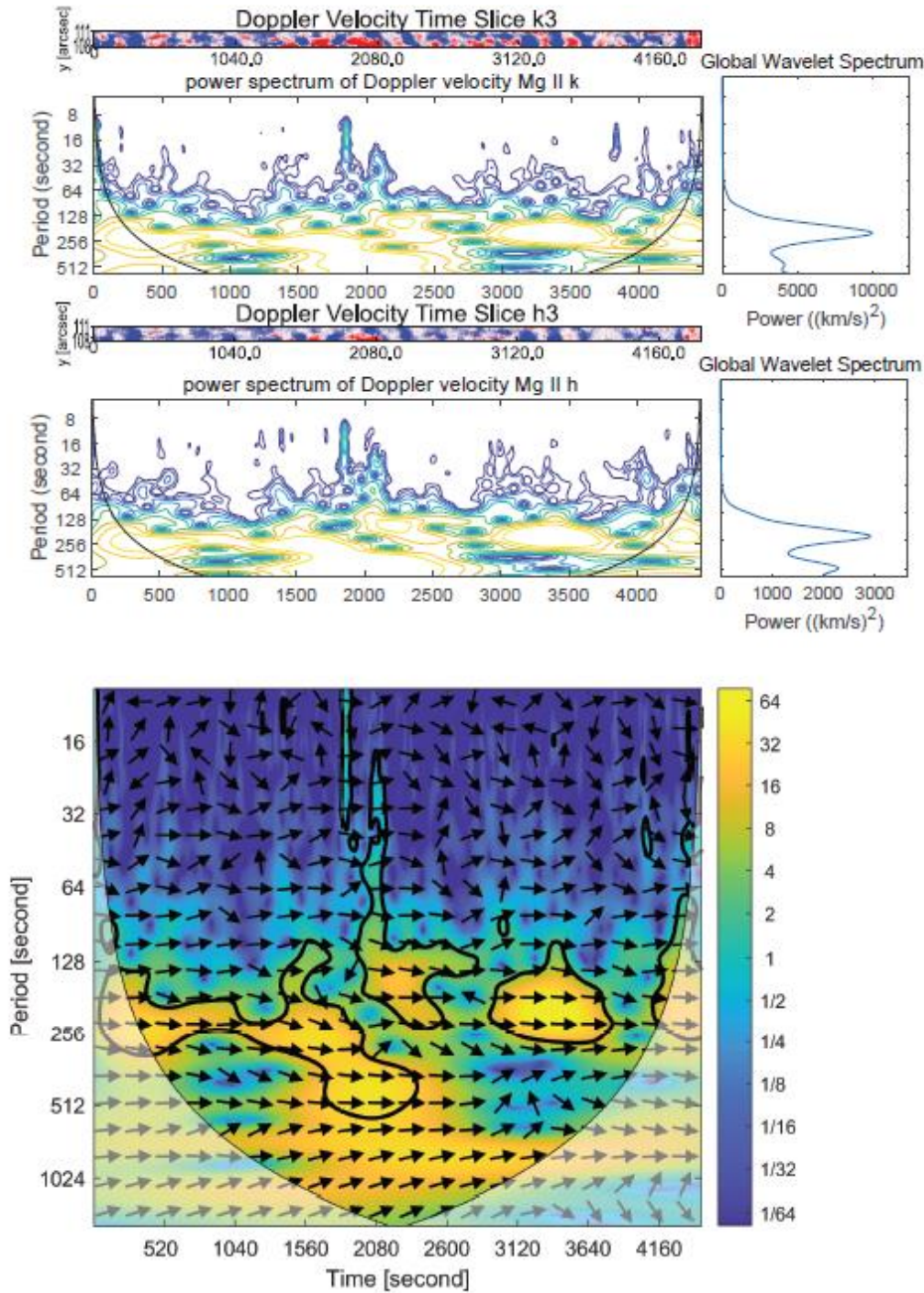

**Figure 7.** Point 1 (P1)-$1^{st}$ row: Doppler velocity time slice diagram at k3 peak of Mg II spectral line core. In this diagram, the speeds are shown in blue, white, and red form, and from -40 km / s to +40 km / s. $2^{nd}$ row-right:global wavelet spectrum of Doppler velocity of $k_3$ peak of Mg II spectral line core. $2^{nd}$ row-left:Doppler velocity wavelet analysis at $k_3$ peak of Mg II spectral line core. values normalized to mean value, with blueish depicting lower values and yellowish depicting higher ones. $3^{rd}$ row: Doppler velocity time slice diagram at $h_3$ peak of Mg II spectral line core. In this diagram, the speeds are shown in blue, white, and red form, and from -40 km / s to +40 km / s. $4^{th}$ row-right:global wavelet spectrum of Doppler velocity of $h_3$ peak of Mg II spectral line core. $4^{th}$ row-left: Doppler velocity wavelet analysis at $h_3$ peak of Mg II spectral line core. values normalized to mean value, with blueish depicting lower values and yellowish depicting higher ones. $5^{th}$ row:cross wavelet analysis of Doppler velocities of $k_3$ and $h_3$ peak of Mg II spectral line core.

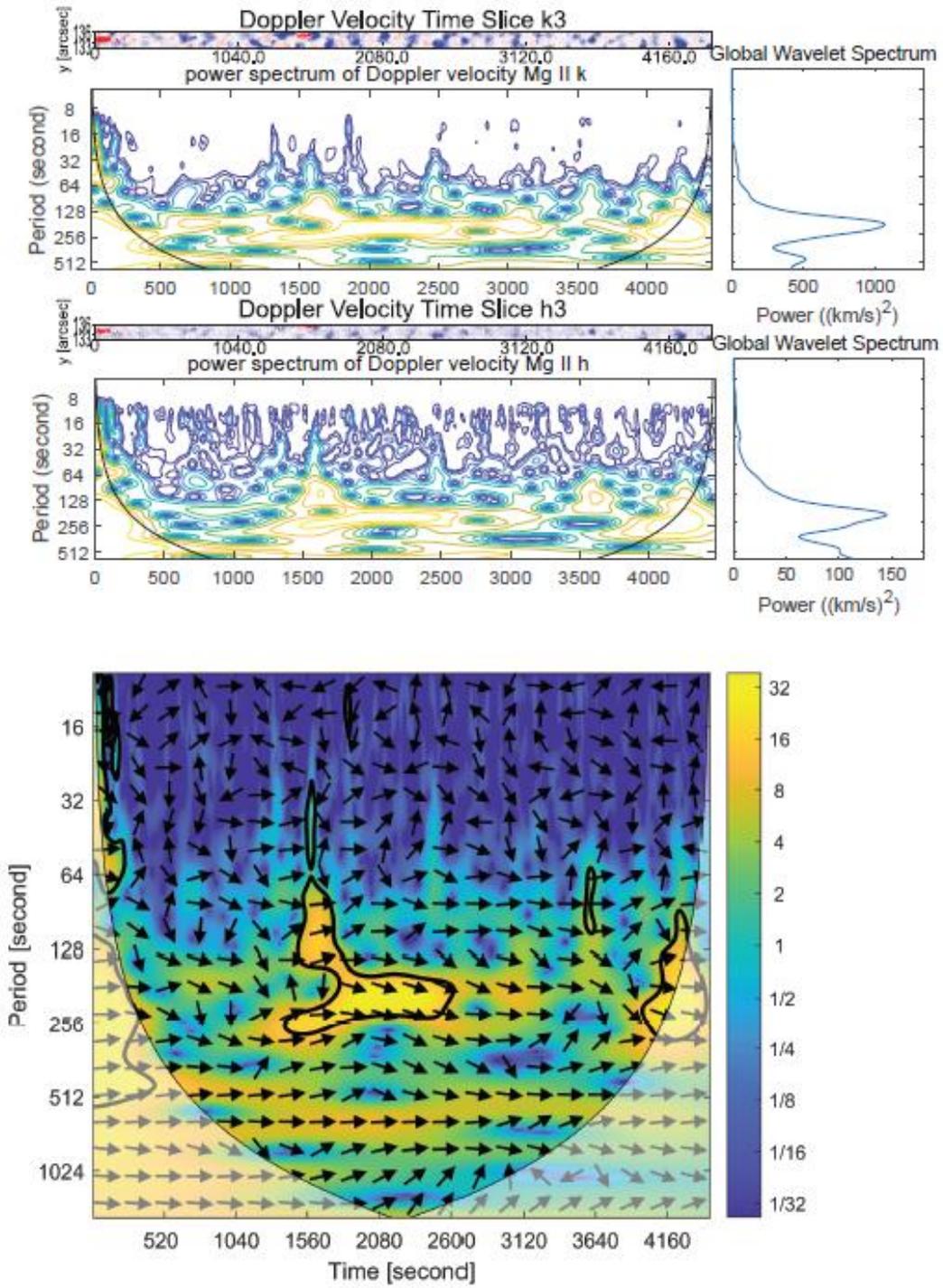

Figure 8. Point 2 (P2)-As figure 6 explanation for P2

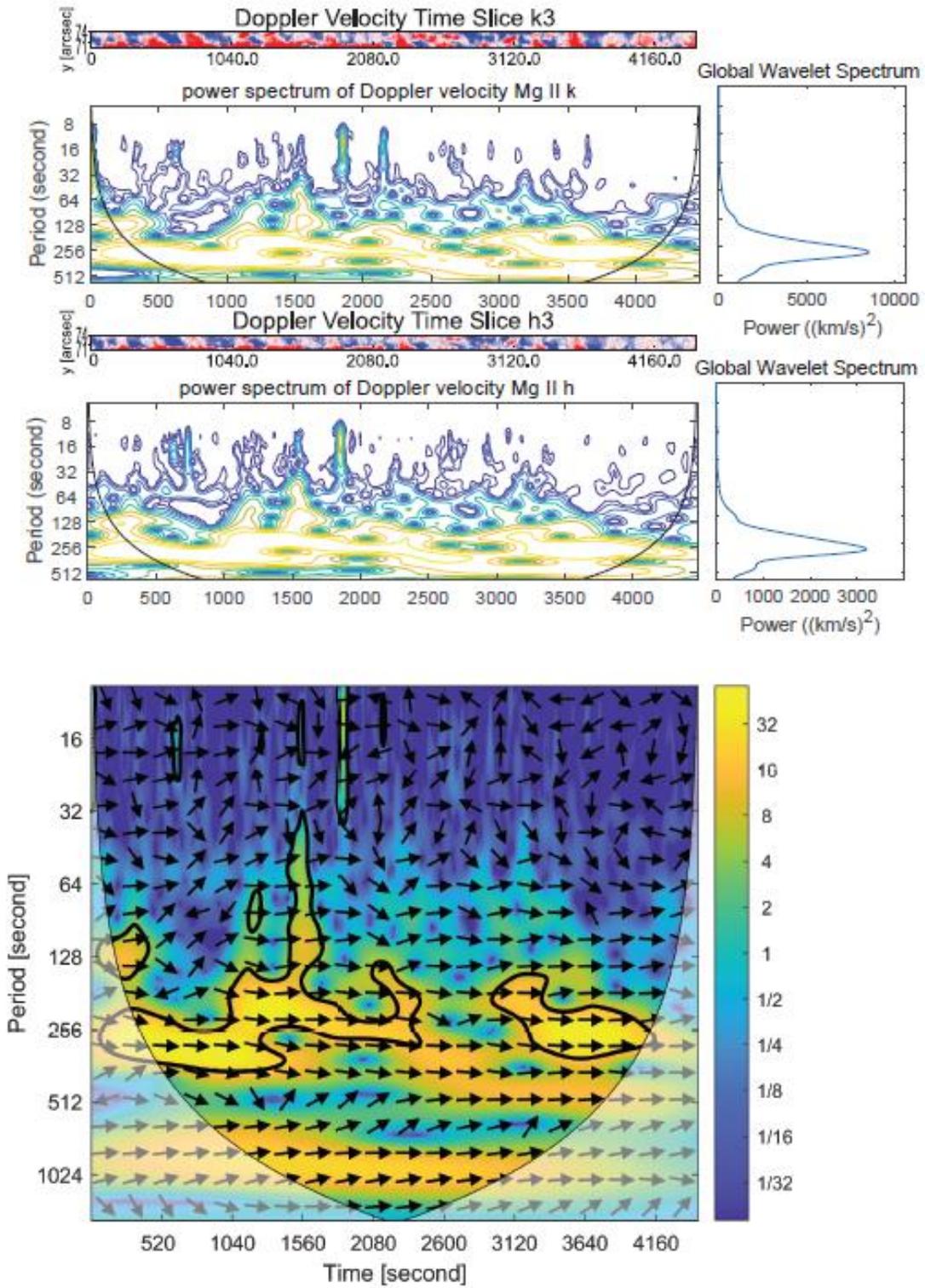

Figure 9. Point 3 (P3)-As figure 6 explanation for P3

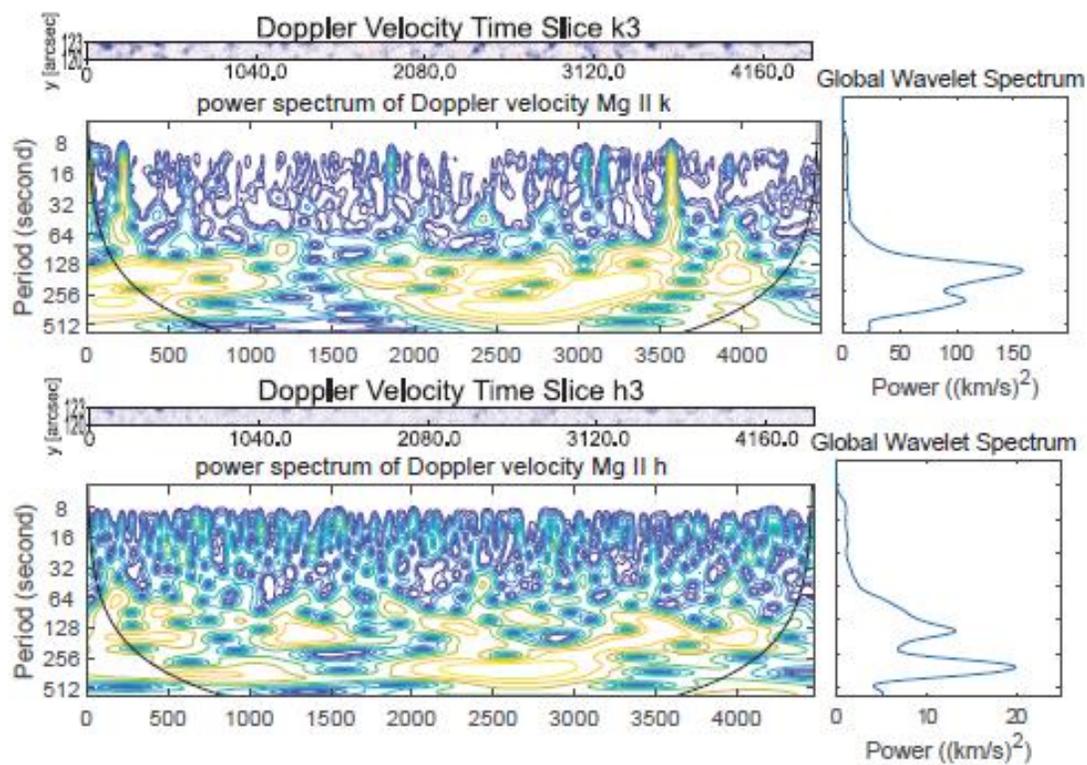

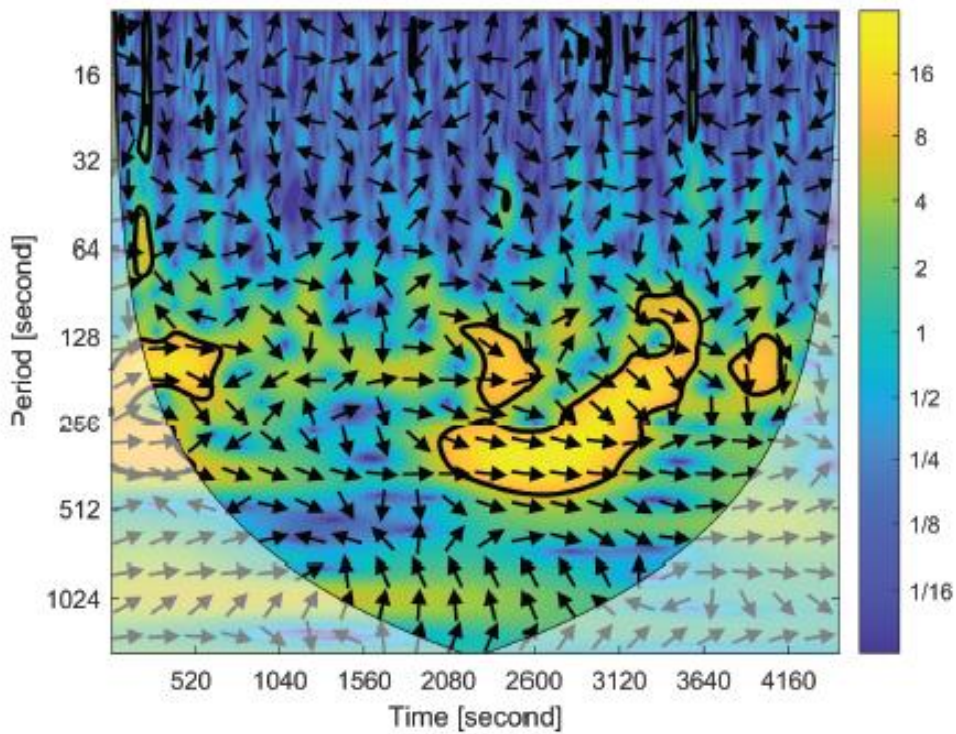

Figure 10. Point 4 (P4)-As figure 6 explanation for P4

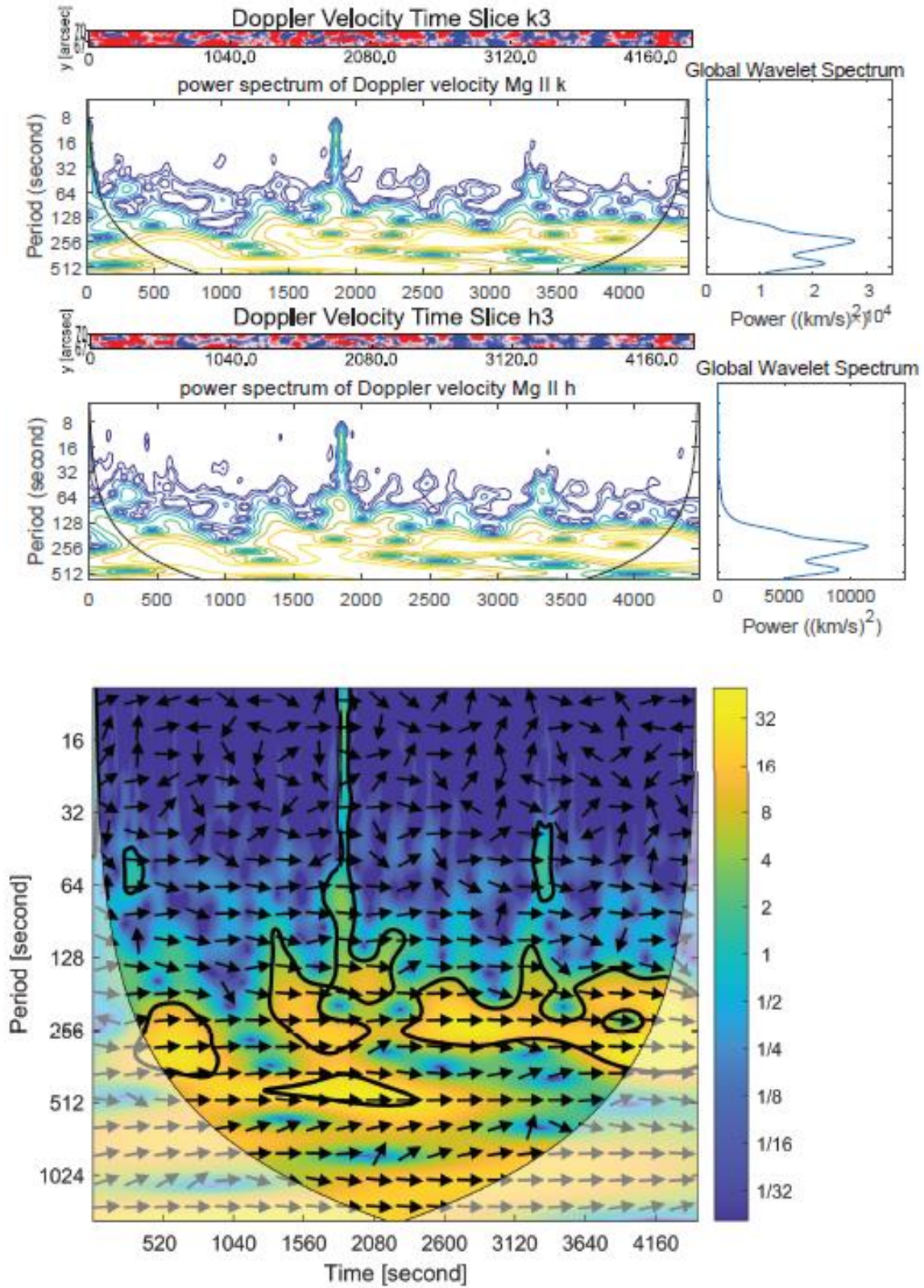

Figure 11. Point 5 (P5)-As figure 6 explanation for P5

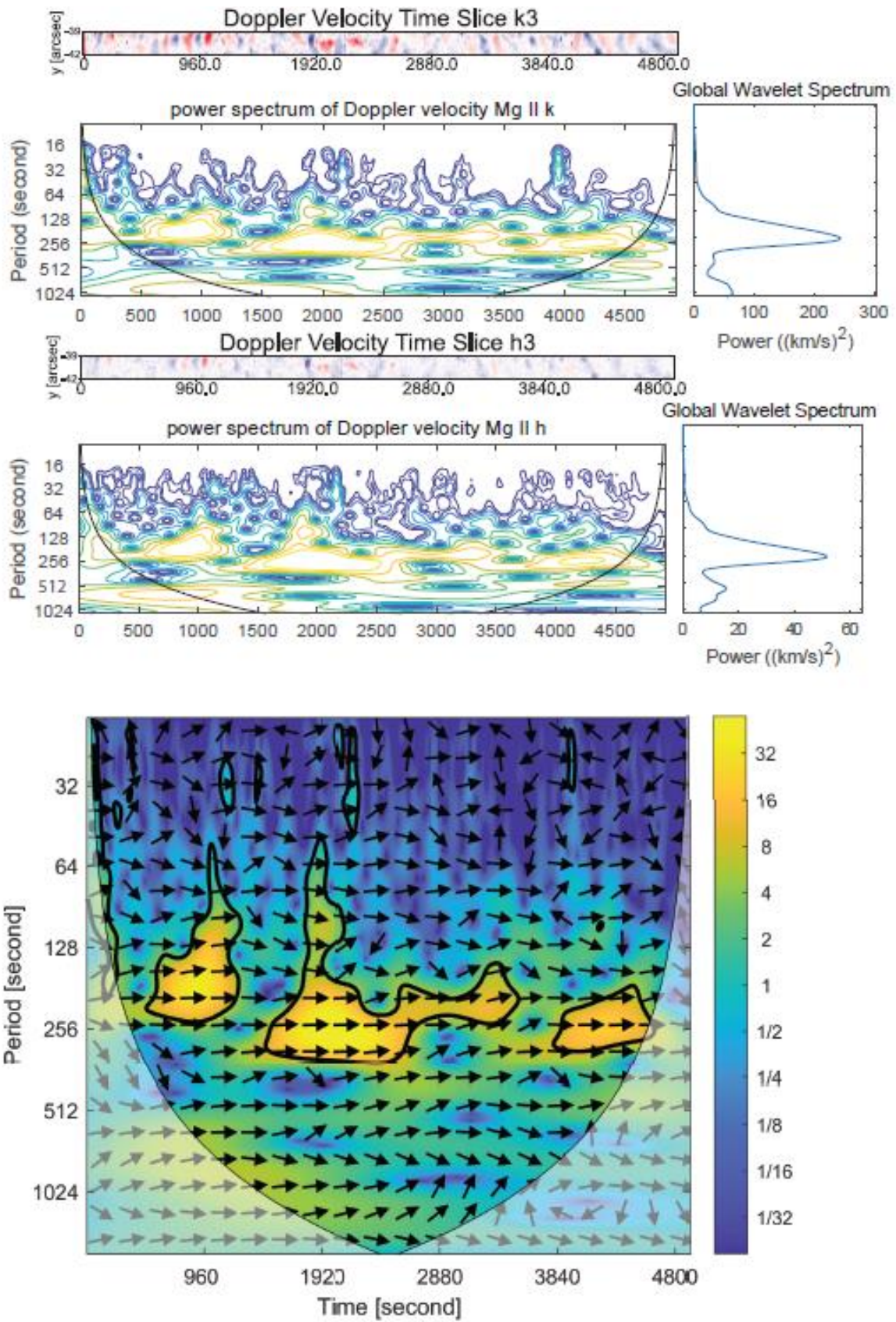

Figure 12. Point 6 (P6)-As figure 6 explanation for P6

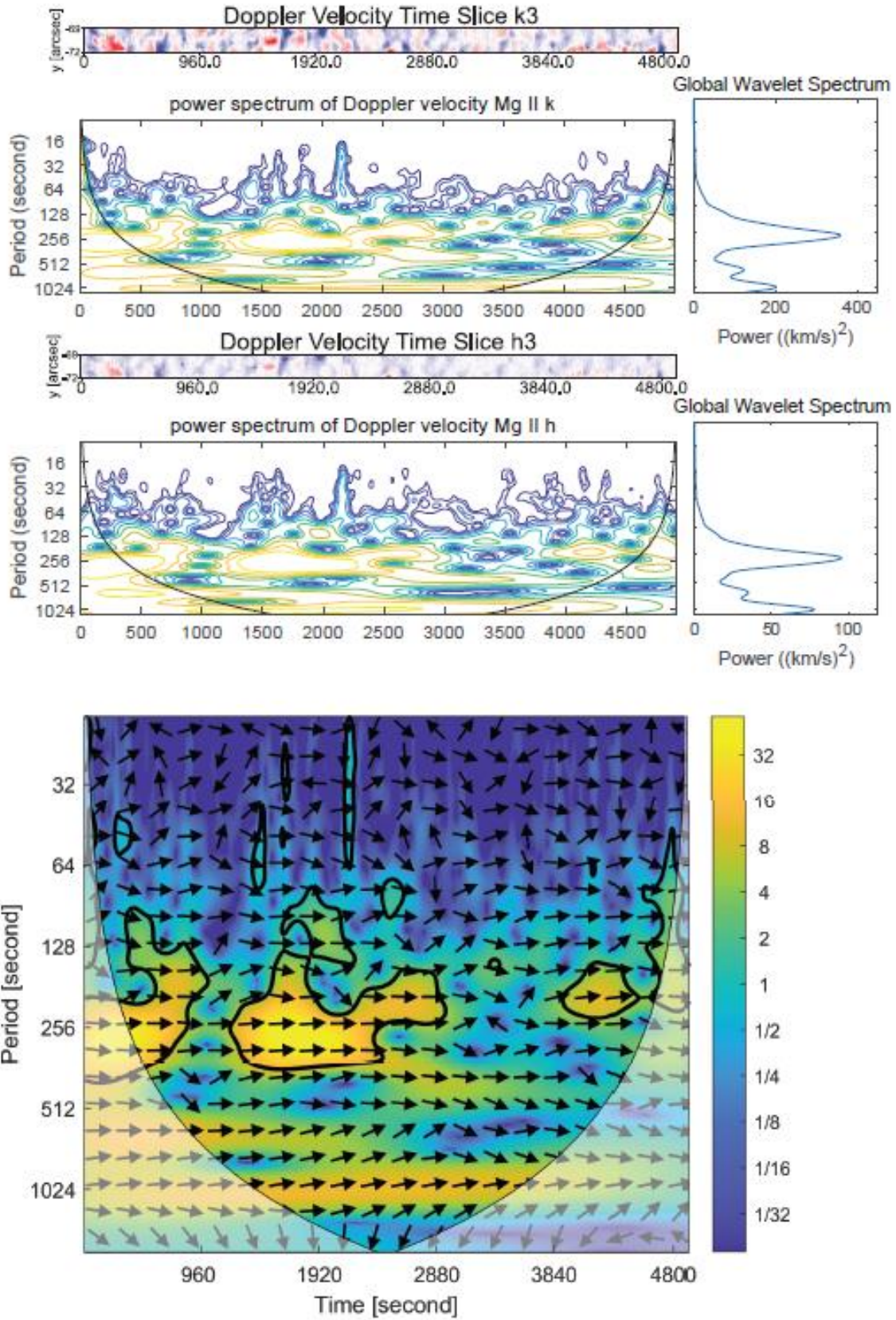

Figure 13. Point 7 (P7)-As figure 6 explanation for P7

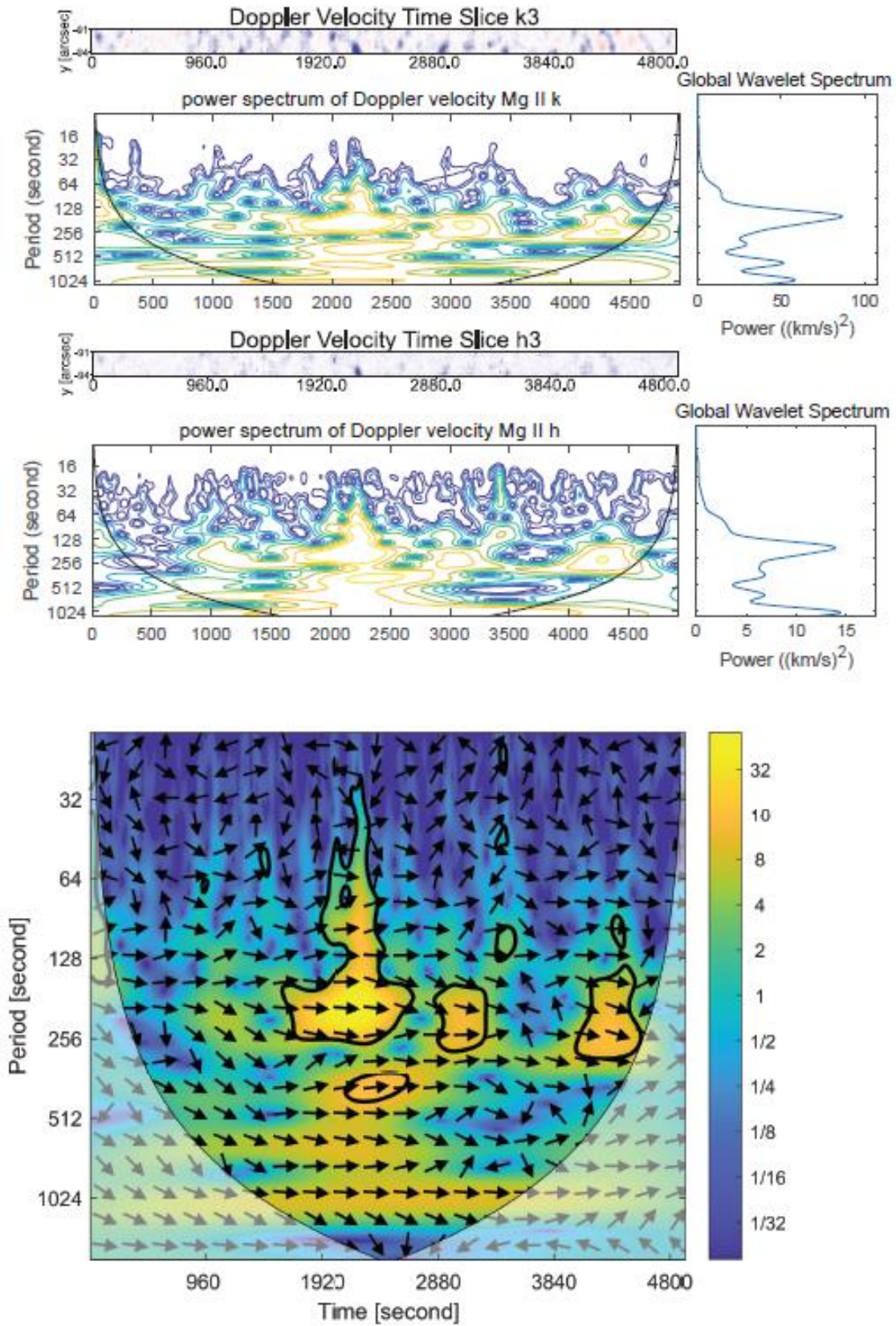

Figure 14. Point 8 (P8)-As figure 6 explanation for P8

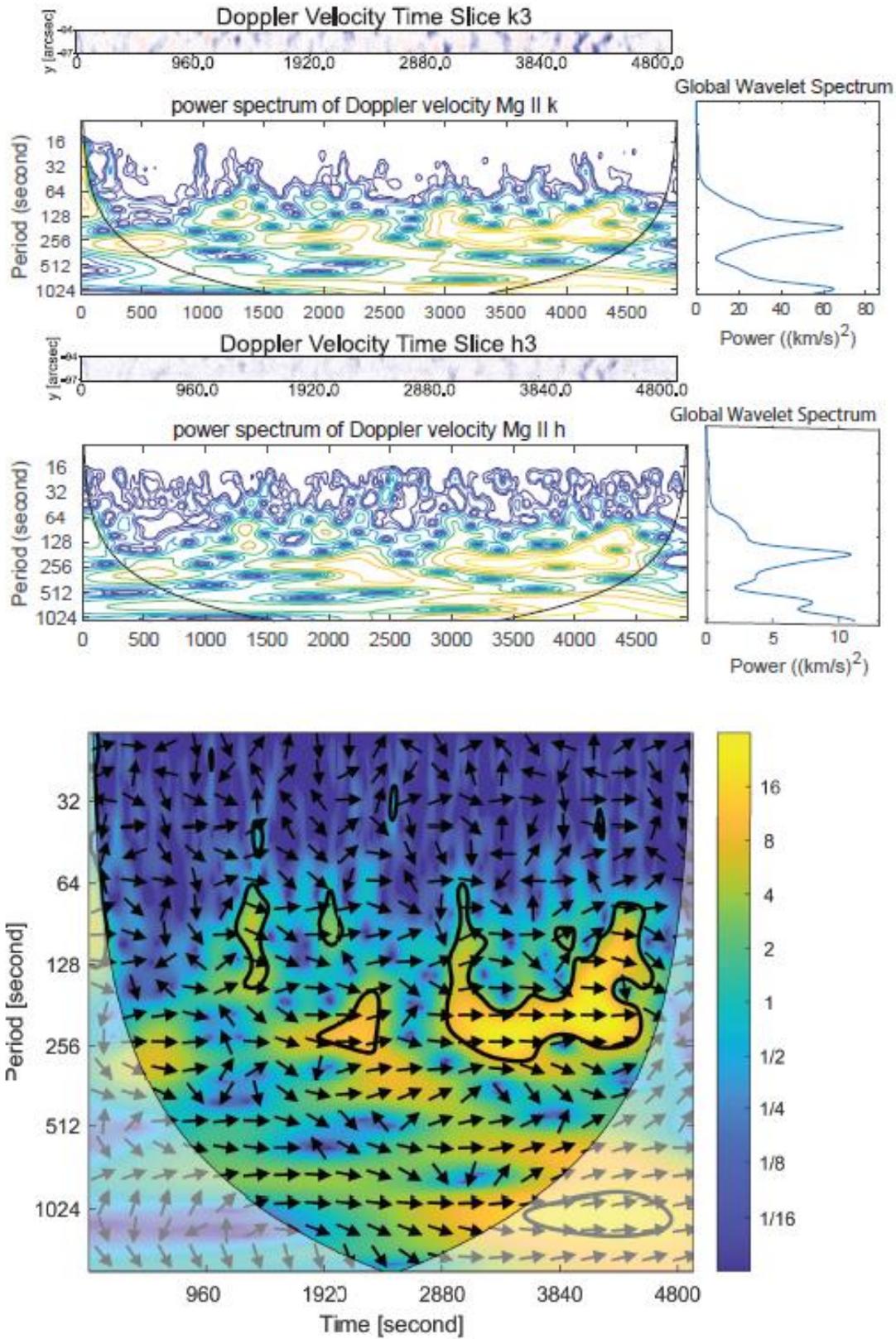

Figure 15. Point 9 (P9)-As figure 6 explanation for P9